\documentclass[12pt]{iopart}

\expandafter\let\csname equation*\endcsname=\relax 
\expandafter\let\csname endequation*\endcsname=\relax 

\usepackage[x11names]{xcolor}

\usepackage{subcaption}

\usepackage[pdftex]{graphicx}
\usepackage{amsmath,amssymb}
\usepackage{bm,caption,setspace,xspace}
\usepackage{color}
\usepackage{comment}

\usepackage[backend=bibtex,style=numeric,autocite=superscript,sorting=none,minbibnames=5,maxbibnames=6,doi=false,isbn=false,url=false]{biblatex}
\bibliography{literature.bib,library.bib}

\renewbibmacro{in:}{}
\AtEveryBibitem{\clearfield{month}}

\graphicspath{{./Figures/}}

\let\vec\bm

\newcommand{\uu}[1]{\ensuremath{\, \mathrm{#1}}} 
\newcommand{\chem}[1]{\ensuremath{\mathrm{#1}}} 

\newcommand{\ra}{\ensuremath{\rightarrow}\xspace}

\newcommand{\DB}[1]{\textcolor{blue}{DB: #1}}
\newcommand{\NLF}[1]{\textcolor{orange}{NLF: #1}}

\newcommand{\DK}[1]{\textcolor{violet}{[Derek: #1]}}
\newcommand{\AW}[1]{\textcolor{purple}{[Arne: #1]}}
\newcommand{\DA}[1]{\textcolor{magenta}{[Deniz: #1]}}

\makeatletter
\let\start@align@nopar\start@align
\let\start@gather@nopar\start@gather
\let\start@multline@nopar\start@multline
\long\def\start@align{\par\start@align@nopar}
\long\def\start@gather{\par\start@gather@nopar}
\long\def\start@multline{\par\start@multline@nopar}
\makeatother

\begin{document}

\title[Quantum sensitivity limits of nuclear magnetic resonance experiments\ldots]{Quantum sensitivity limits of nuclear magnetic resonance experiments searching for new fundamental physics}



\author{Deniz~Aybas}
\address{Department of Physics, Boston University, Boston, MA 02215, USA}
\address{Department of Electrical and Computer Engineering, Boston University, Boston, MA 02215, USA}
\author{Hendrik~Bekker}
\address{Johannes Gutenberg-Universit{\"a}t Mainz, 55128 Mainz, Germany}
\author{John~W.~Blanchard}
\address{Helmholtz-Institut, GSI Helmholtzzentrum f{\"u}r Schwerionenforschung, 55128 Mainz, Germany}
\author{Dmitry~Budker}
\address{Helmholtz-Institut, GSI Helmholtzzentrum f{\"u}r Schwerionenforschung, 55128 Mainz, Germany}
\address{Johannes Gutenberg-Universit{\"a}t Mainz, 55128 Mainz, Germany}
\address{Department of Physics, University of California, Berkeley, California 94720-7300, USA}
\author{Gary~P.~Centers}
\address{Helmholtz-Institut, GSI Helmholtzzentrum f{\"u}r Schwerionenforschung, 55128 Mainz, Germany}
\address{Johannes Gutenberg-Universit{\"a}t Mainz, 55128 Mainz, Germany}
\author{Nataniel~L.~Figueroa}
\address{Helmholtz-Institut, GSI Helmholtzzentrum f{\"u}r Schwerionenforschung, 55128 Mainz, Germany}
\address{Johannes Gutenberg-Universit{\"a}t Mainz, 55128 Mainz, Germany}
\author{Alexander~V.~Gramolin}
\address{Department of Physics, Boston University, Boston, MA 02215, USA}
\author{Derek~F.~Jackson~Kimball}
\address{Department of Physics, California State University - East Bay, Hayward, California 94542-3084, USA}
\author{Arne~Wickenbrock}
\address{Helmholtz-Institut, GSI Helmholtzzentrum f{\"u}r Schwerionenforschung, 55128 Mainz, Germany}
\address{Johannes Gutenberg-Universit{\"a}t Mainz, 55128 Mainz, Germany}
\author{Alexander~O.~Sushkov\footnote{Electronic address: asu@bu.edu}}
\address{Department of Physics, Boston University, Boston, MA 02215, USA}
\address{Department of Electrical and Computer Engineering, Boston University, Boston, MA 02215, USA}
\address{Photonics Center, Boston University, Boston, MA 02215, USA}

\date{\today}

\baselineskip16pt

\maketitle

\begin{abstract}
Nuclear magnetic resonance is a promising experimental approach to search for ultra-light axion-like dark matter. Searches such as the  cosmic axion spin-precession experiments (CASPEr) are ultimately limited by quantum-mechanical noise sources, in particular, spin-projection noise. 
We discuss how such fundamental limits can potentially be reached. We consider a circuit model of a magnetic resonance experiment and quantify three noise sources: spin-projection noise, thermal noise, and amplifier noise. Calculation of the total noise spectrum takes into account the modification of the circuit impedance by the presence of nuclear spins, as well as the circuit back-action on the spin ensemble. Suppression of the circuit back-action is especially important in order for the spin-projection noise limits of searches for axion-like dark matter to reach the quantum chromodynamic axion sensitivity.
\end{abstract}

\section{Overview of fundamental physics measurements using magnetic resonance}

Nuclear magnetic resonance (NMR) experiments have long been at the forefront of precision tests of fundamental physics \cite{safronova2018search,demille2017probing}. One of the earliest such efforts was the neutron-beam NMR experiment carried out in the 1950s by Purcell, Ramsey, and Smith \cite{Pur50,smith1957experimental} to search for a parity (P) and time-reversal (T) violating permanent electric dipole moment (EDM) of the neutron. It is interesting to note that this earliest EDM experiment focused on the P-violating character of the EDM -- only after the discovery of P-violation in $\beta$-decay \cite{wu1957experimental} did Landau point out that EDMs are also T-violating \cite{Lan57}. Modern EDM experiments \cite{regan2002new,hudson2011improved,baron2014order,graner2016reduced,cairncross2017precision,andreev2018improved,allmendinger2019measurement,abel2020measurement}, many employing NMR methods, are motivated by the fact that an EDM with a magnitude measurable with present techniques would be evidence of a new source of CP-violation (where C represents charge conjugation); additional sources of CP-violation beyond those in the Standard Model are needed to explain the cosmological asymmetry between matter and antimatter \cite{khriplovich2012cp}. An experiment to test the isotropy of space inspired by Mach's principle and using NMR was carried out by Hughes, Robinson, and Beltran-Lopez in 1960 \cite{hughes1960upper}; since then there have been many related NMR experiments testing Lorentz invariance (see, for example, Refs.~\cite{brown2010new,smiciklas2011new,gemmel2010limit}). Another early use of NMR methods to test fundamental physics was in a series of experiments searching for couplings between intrinsic spin and gravity \cite{Vas69,You69,Win72,Win91,Ven92}, and experiments along these lines are still being actively pursued \cite{kimball2017constraints,lee2018improved,wang2020single,fadeev2020gravity}. The existence of ``new'' spin-0 or spin-1 bosons may imply the existence of exotic spin-dependent interactions \cite{Moo84,Dob06,fadeev2019revisiting}, which can also be searched for using NMR. For example, Ramsey searched for exotic spin-dependent couplings between protons using NMR measurements \cite{Ram79}, and here too, recent experiments build on Ramsey's early work \cite{Vas09,Hun13,Led13}.

The tools of NMR can also be used to search for ultralight bosonic dark matter \cite{Gra13,Bud14}, in particular axions and axion-like particles (ALPs) \cite{Gra15review}. If dark matter consists primarily of particles with mass $m_a \lesssim 1~{\rm eV}/c^2$, their density must be so large that, rather than individual particles, their behavior can be treated as a highly coherent classical field oscillating near the Compton frequency $\omega_1 = m_a c^2 / \hbar$. This axion-like field can both generate oscillating nuclear EDMs via a coupling to gluons and act as an oscillating ``pseudo-magnetic'' field via a coupling of the gradient of the axion-like field to fermion spins \cite{Gra13,Bud14}. In either case, the resultant interaction between axion dark matter fields and nuclear spins is similar to that of an oscillating magnetic field, and thus can be searched for using the tools of NMR. This is the central concept of the Cosmic Axion Spin Precession Experiment (CASPEr) \cite{Bud14,garcon2017cosmic,wu2019search,kimball2020overview,garcon2019constraints,Aybas2021}.

\section{Basics of NMR}

\begin{figure}[!ht]
\begin{center}
\includegraphics[width=0.5\textwidth]{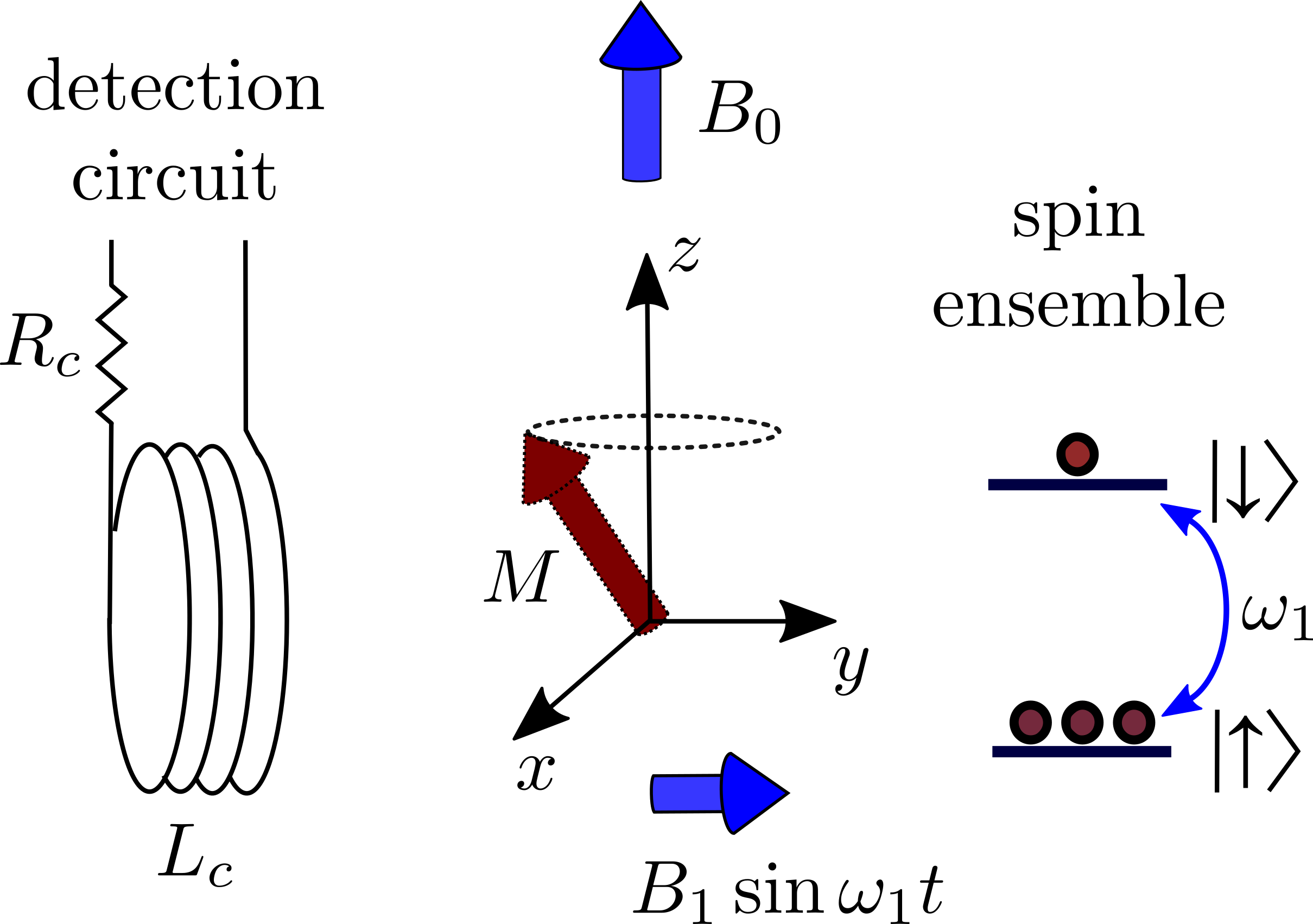}
\end{center}
\newpage
\caption{\linespread{1.0}\fontsize{9}{12}\selectfont
A schematic of a typical NMR experiment. $M$ is the nuclear spin magnetization of the sample. $B_0$ is the bias magnetic field, and $B_1\sin{\omega_1t}$ is either an externally-applied excitation field, or the ``pseudo-magnetic'' field due to interaction with ultralight dark matter. The spin-1/2 level diagram indicates spin polarization as larger population in the ground spin sublevel, and spin coherence induced by the excitation field $B_1$, if it is resonant with the spin Larmor frequency.
The inductive detection circuit includes a coil with inductance $L_c$ and resistance $R_c$. 
}
\label{fig:NMR_expt_schematic}
\end{figure}
Magnetic resonance encompasses a broad and versatile set of techniques that have found application in a wide range of disciplines.
A typical  NMR experiment investigates nuclear spin dynamics in an applied bias magnetic field (Fig.\,\ref{fig:NMR_expt_schematic}), although in zero- and ultralow-field (ZULF) NMR the bias field may be small or absent~\cite{Blanchard2016emagres}. In pulsed magnetic resonance experiments, the spins are excited with a sequence of resonant radiofrequency (RF) pulses, and the subsequent spin evolution is detected. In the context of fundamental physics, searches for permanent electric dipole moments usually employ the pulsed scheme~\cite{abel2020measurement}. 
In continuous wave (CW) magnetic resonance experiments, the excitation field is present continuously. Spin-based dark matter haloscope experiments usually employ the CW scheme~\cite{Graham2018_spin_prec}. Here we will focus on CW magnetic resonance, which is also convenient for considering the problem of spin-projection noise. In our treatment we will neglect saturation effects, working in the limit of weak drive fields.


\subsection{Nuclear spin susceptibility}

It is convenient to quantify the CW NMR response of a spin ensemble by its frequency-dependent magnetic susceptibility~\cite{Slichter1978}. In the limit of a weak drive, the complex susceptibility $\chi = \chi'-i\chi''$ is given by:
\begin{align}
\begin{split}
\chi'(\omega) & = \frac{1}{2}\chi_0\omega_0T_2^*\frac{(\omega_0-\omega)T_2^*}{1+(\omega_0-\omega)^2T_2^{*2}},\\
\chi''(\omega) & = \frac{1}{2}\chi_0\omega_0T_2^*\frac{1}{1+(\omega_0-\omega)^2T_2^{*2}},
\end{split}
\label{eq:110}
\end{align}
where $\omega_0=\gamma B_0$ is the resonance frequency of spins with gyromagnetic ratio $\gamma$ in a bias magnetic field $B_0$, susceptibility $\chi_0$ is defined via the sample magnetization: $\chi_0B_0=\mu_0M_0$, and $T_2^*$ is the transverse relaxation time, including damping due to the interaction with the pickup circuit. 
We use SI units, with permeability of free space $\mu_0$.

Let us re-write the above in terms of magnetization:
\begin{align}
\begin{split}
\chi' & = \frac{1}{2}\gamma (\mu_0M_0)T_2^*\frac{(\omega_0-\omega)T_2^*}{1+(\omega_0-\omega)^2T_2^{*2}},\\
\chi'' & = \frac{1}{2}\gamma (\mu_0M_0)T_2^*\frac{1}{1+(\omega_0-\omega)^2T_2^{*2}}.
\end{split}
\label{eq:120}
\end{align}
Assuming that spin polarization is much less than unity, we can connect magnetization with spin temperature $\theta_s$:
\begin{align}
M_0=\frac{n\hbar^2\gamma^2I(I+1)B_0}{3k_B\theta_s},
\label{eq:M_0}
\end{align}
where $n$ is spin number density, $I$ is the nuclear spin, $\hbar$ is the reduced Planck constant, and $k_B$ is the Boltzmann constant. In thermal equilibrium, $\theta_s=\theta_c$, where $\theta_c$ is the physical system temperature. If spins are hyperpolarized then $\theta_s<\theta_c$, and if they have been saturated, then $\theta_s>\theta_c$. 
We note that negative spin temperatures can, of course, also be achieved, in which case the magnetization is also negative, corresponding to a population excess in the spin sub-level with higher energy.

\subsection{CW NMR signal}

In the limit of a weak continuous coherent drive at frequency $\omega$, the spin response can be quantified by the unsaturated steady-state transverse magnetization $M_1$, calculated using the susceptibility in Eq.\,(\ref{eq:120}):
\begin{align}
\mu_0M_1=\frac{\mu_0M_0\Omega_1 T_2^*}{1+(\omega_0-\omega)^2T_2^{*2}},
\label{eq:140}
\end{align}
where $\Omega_1$ is the drive Rabi frequency, proportional to the coupling strength of a beyond-Standard-Model field that the experiment is designed to search for.

\subsection{NMR detection circuit coupled to a spin ensemble}\label{sec:2.3}


\noindent
There are many experimental approaches designed to detect NMR. The most basic approach is to use a pickup coil coupled to the spin ensemble. The transverse magnetization precesses around the leading field at the Larmor frequency, creating an oscillating magnetic flux, which induces a Faraday voltage across the coil, Fig.~\ref{fig:NMR_expt_schematic}. As discussed below, a resonant circuit is often used to couple this voltage to a sensitive amplifier. However at first we focus on the basic elements of the inductive detection scheme: the pickup coil inductance $L_c$ and the series resistance $R_c$. For an empty solenoid coil the inductance is given by $L_c = \mu_0\eta^2A/l$, where $\eta$ is the number of turns, $A$ is the coil area, and $l$ is its length. The resistance determines the coil quality factor: $Q_c=\omega L_c/R_c$.

The magnetic permeability of the spin sample changes the coil inductance and resistance when the spin sample is inserted. With the spin sample in place, the pickup coil impedance becomes
\begin{align}
Z = R_c + i\omega L_c(1+q\chi) = (R_c+q\omega L_c\chi'') + i\omega L_c(1+q\chi'),
\label{eq:220}
\end{align}
where $q$ is a filling factor. We define the spin resistance $R_s$ and spin inductance $L_s$, such that 
$Z = (R_c+R_s) + i\omega (L_c+L_s)$, with
\begin{align}
L_s &= \frac{q}{2} L_c \gamma (\mu_0M_0)T_2^*\frac{(\omega_0-\omega)T_2^*}{1+(\omega_0-\omega)^2T_2^{*2}},\\
R_s &= \frac{q}{2}\omega L_c\gamma (\mu_0M_0)T_2^*\frac{1}{1+(\omega_0-\omega)^2T_2^{*2}}.
\label{eq:R_s}
\end{align}
Thus, through Eqs.~(\ref{eq:220}-\ref{eq:R_s}), it is seen that the spins in the sample modify the impedance of the circuit.

The circuit, in turn, modifies the properties of the spin ensemble via the back-action mechanism, which has historically been called ``radiation damping'' (although there is no radiation involved)~\cite{Hoult2001FID}. The current in the pickup circuit creates a magnetic field, which resonantly couples back to the spin ensemble. This circuit back-action can be described as a spin-relaxation mechanism, with the rate given by~\cite{Augustine2002b}
\begin{align}
\frac{1}{T_r} = \frac{1}{2}qQ_c\gamma\mu_0M_0.
\label{eq:113}
\end{align}
If the intrinsic coherence time of the spin ensemble is $T'_2$, then when the spins are coupled to the pickup circuit, their coherence time becomes
\begin{align}
\frac{1}{T_2^*} = \frac{1}{T'_2} + \frac{1}{T_r} = \frac{1}{T'_2} + \frac{qQ_c\gamma\mu_0M_0}{2}.
\label{eq:115}
\end{align}

The circuit back-action is a well-known phenomenon in the field of NMR, and several ways to suppress it have been developed~\cite{Broekaert1995,Maas1995}.
The most promising approach for a precision fundamental physics experiment is likely to implement a feedback scheme that cancels the current in the pickup coil, induced by precessing spin magnetization. Such a scheme is commonly used with SQUID sensors in order to avoid cross-talk~\cite{Hatridge2011_Dispersive}.

\section{Spin-projection noise}
\label{sec:SPN}

\subsection{The standard quantum limit}\label{sec:SQL}

Spin-projection noise is closely related to the Heisenberg uncertainty principle and the standard quantum limit. A simple way to understand spin-projection noise is to consider the following thought experiment. Suppose a single spin-1/2 is prepared in the ``spin-up'' state, namely the quantum state with the $s_z=1/2$ spin projection along the z axis (Fig.\,\ref{fig:NMR_expt_schematic}). Then a measurement of the $s_y$ spin component is performed. There are two possible outcomes: $s_y=+1/2,-1/2$, and they are equally likely. If this sequence is repeated $N$ times, or the experiment is performed on $N$ uncorrelated spins, 
then the (random) mean value of the $s_y$ spin component is normally distributed, with standard deviation $\sqrt{N}/2$. The uncertainty in the transverse spin projection corresponds to a $\delta\theta \approx 1/\sqrt{N}$ uncertainty in the polar angle of the spin, or the spin ensemble. This uncertainty is the origin of the spin-projection noise.

In an NMR experiment, a continuous measurement of one of the components of the transverse magnetization $M_y = (\hbar\gamma/V)\sum s_y$ is usually performed, here $V=qAl$
is the sample volume.
In an applied bias magnetic field, $M_y$ oscillates at the Larmor angular frequency $\omega_0=\gamma B_0$. The coherence time of $M_y$ is $\approx T_2^*$ and the root-mean-squared magnetization due to the spin-projection noise is
\begin{align}
\sqrt{\langle M^2_y\rangle}\approx\frac{\hbar\gamma}{V}\sqrt{N}.
\label{eq:315}
\end{align}
We have dropped factors of order unity, which will be tracked more carefully in the subsequent sections.
Let us denote the frequency-domain amplitude spectral density of $M_y$ by $\tilde{M}_y$ (here and below we mark the angular frequency-domain Fourier-transformed quantities by a tilde). It has a peak centered at $\omega_0$, with FWHM linewidth $\Delta \omega = 2/T_2^*$ [see Eq.\eqref{eq:140}] and amplitude
\begin{align}
\tilde{M}_y(\omega=\omega_0)\approx\frac{\hbar\gamma}{V}\sqrt{N}\sqrt{T_2^*}.
\label{eq:310}
\end{align} 
We note that the root-mean-squared magnetization noise in Eq.~\eqref{eq:315} is the square root of the area under the power spectrum [Eq.~\eqref{eq:310} squared and multiplied by $\Delta\omega$].

\subsection{The circuit model of spin-projection noise}\label{sec:circuitmodel}
An alternative way to derive the spin-projection noise \eqref{eq:310} is to consider the Nyquist noise generated by the spin-induced resistance $R_s$. According to the fluctuation-dissipation theorem~\cite{Kubo_1966_FDT}, the power spectral density of the voltage noise created by the spin ensemble at spin temperature $\theta_s$ is
\begin{align}
\tilde{V}^2_s(\omega) = \frac{2R_s}{\pi}\frac{\hbar\omega}{2}\coth\frac{\hbar\omega}{2k_B\theta_s},
\label{eq:V_s^2(omega)}
\end{align}
In the classical regime $\hbar\omega\ll k_B\theta_s$, and this simplifies to the well-known formula $\tilde{V}^2_s = 2R_sk_B\theta_s/\pi$. Using Eqs.\,(\ref{eq:M_0}) and (\ref{eq:R_s}), we find the Nyquist noise spectrum:
\begin{align}
\tilde{V}^2_s(\omega) = \frac{q}{2\pi}\mu_0\hbar^2\gamma^2n\omega^2L_c T_2^*\frac{1}{1+(\omega_0-\omega)^2T_2^{*2}}.
\label{eq:330}
\end{align}
Note that the spin-noise voltage is independent of spin temperature in both the classical regime $\hbar\omega\ll k_B\theta_s$ [considered above, here $\theta_s$ from the argument of cotangent cancels with the $\theta_s$ in the denominator of $M_0$, see Eqs.\,\eqref{eq:M_0} and \eqref{eq:R_s}] and the quantum regime $\hbar\omega\gg k_B\theta_s$ [where $\coth(\hbar\omega/2k_B\theta_s)\rightarrow 1$ and spin polarization approaches unity].

We note the equivalence between the standard quantum limit treatment and the Nyquist treatment. 
For inductive detection, we can use the transfer coefficient $\alpha_0$ to convert transverse magnetization $M_1$ to voltage $V_1$ across the pickup coil: $V_1=\alpha_0\mu_0M_1$~\cite{Aybas2021}. Faraday's law of induction gives
\begin{align}
|\alpha_0| = q\omega\eta A,
\label{eq:333}
\end{align}
where $\omega$ is the signal frequency.
Therefore the oscillating magnetization (\ref{eq:310}) creates an induced voltage in the coil:
\begin{align}
\tilde{V}_s(\omega = \omega_0) = q\omega_0\eta A\mu_0\frac{\hbar\gamma}{V}\sqrt{N}\sqrt{T_2^*}.
\label{eq:340}
\end{align}
We can verify that this reproduces (up to a numerical factor) the on-resonance voltage in Eq.~(\ref{eq:330}) by squaring and substituting $N=nV=nqAl$ and $L_c=\mu_0\eta^2A/l$.

\section{Observations of spin-projection noise and its applications}
Already in 1946, Bloch noted that spin-projection noise should be observable using NMR~\cite{Bloch1946Induction}. Namely, a sample of $N$ spins of magnetic moment $\mu$ are statistically highly unlikely to perfectly cancel, so that the sample has an instantaneous nonzero magnetic moment of the order $(N)^{1/2} \mu$. Along the same lines, the instantaneous magnetic moment of a sample in a magnetic field pointing along the $z$-axis has a nonzero component perpendicular to $z$ as discussed in Sec.\,\ref{sec:SQL}. Even so, the time average of this transverse magnetic moment will converge to zero with the characteristic time constant $T_2^*$. If other noise sources are sufficiently suppressed, the spontaneously fluctuating transverse magnetic moment can be measured. This was first achieved by Sleator and co-workers for a solid sample containing $^{35}$Cl at liquid helium temperatures~\cite{Sleator1985Spinnoise, Sleator1987Spontaneous}, and later by McCoy and coworkers as well as Gu\'{e}ron and coworkers for liquid-state protons at room temperature~\cite{McCoy1989SpinnoiseRT, gueron_nmr_1989}.

A crucial issue for understanding the role of spin noise in NMR measurements is the role of the detection circuit, which has long been a controversial topic. Based on ideas by Purcell~\cite{purcell1946}, Sleator \textit{et al.} claimed that the low spontaneous emission rate was enhanced due to the increased density of the radiation field in the cavity formed by the resonant high-Q probe circuit~\cite{Sleator1985Spinnoise, Sleator1987Spontaneous}. This interpretation, however, ignores that the sample-coil interaction takes place in the regime of near-field inductive coupling~\cite{Hoult1997}. Hoult and Ginsberg provide a concise historical overview of the arguments, followed by measurements that show that even with a low-Q probe system, spin noise can clearly be observed~\cite{Hoult2001FID}. Thereby, the notion of cavity enhanced emission was rejected and instead one can invoke the concept of virtual photons for a complete quantum electrodynamics (QED) description of the interaction between spins and pickup coil such as done by Engelke~\cite{engelkeVirtualPhotonsMagnetic2010}.

Practically, it is convenient to treat the quantum mechanical spin noise of a macroscopic sample as the Johnson-Nyquist noise of a resistance as described in section~\ref{sec:circuitmodel} and also already applied by, for example, Sleator and co-workers~\cite{Sleator1985Spinnoise, Sleator1987Spontaneous}. However, care needs to be taken here to distinguish between pure spin-projection noise and absorbed circuit noise (ACN) as made explicit in a series of papers by M\"{u}ller, Jerschow, and coworkers~\cite{Giraudeau2010HyperpolNOise, schlagnitweitObservationNMRNoise2010,Muller2013,Bendet-Taicher2014}. The observed noise lineshapes often contain contributions from both effects and are, for example, dominated by ACN when the spin temperature is significantly below that of the detection circuit~\cite{nausnerNonlinearityFrequencyShifts2009}. In that case, a dip in the power spectrum is observed due to energy transfer to the spin sample from the circuit. This improved understanding together with technological advances allowed for experiments optimized for the detection of spin noise, such as at the spin-noise tuning optimum (SNTO)~\cite{Jerschow2010SpinNoise, ferrandNuclearSpinNoise2015a}.
Applications include detection of low $^{13}$C spin concentrations~\cite{schlagnitweitFirstObservationCarbon132012}, monitoring of hyperpolarization without destroying the magnetization~\cite{Desvaux2009}, and imaging~\cite{Muller2006a, ginthorNuclearSpinNoise2020}. These can be considered to be part of a much wider field of spin-noise spectroscopy, of which magneto-optical-rotation spectroscopy is one widely employed technique~\cite{zapasskii2013spin, Sinitsyn2016}.

\section{Technical requirements for performing a spin-projection noise-limited dark matter search}

\begin{figure}[!ht]
\begin{center}
\includegraphics[width=0.8\textwidth]{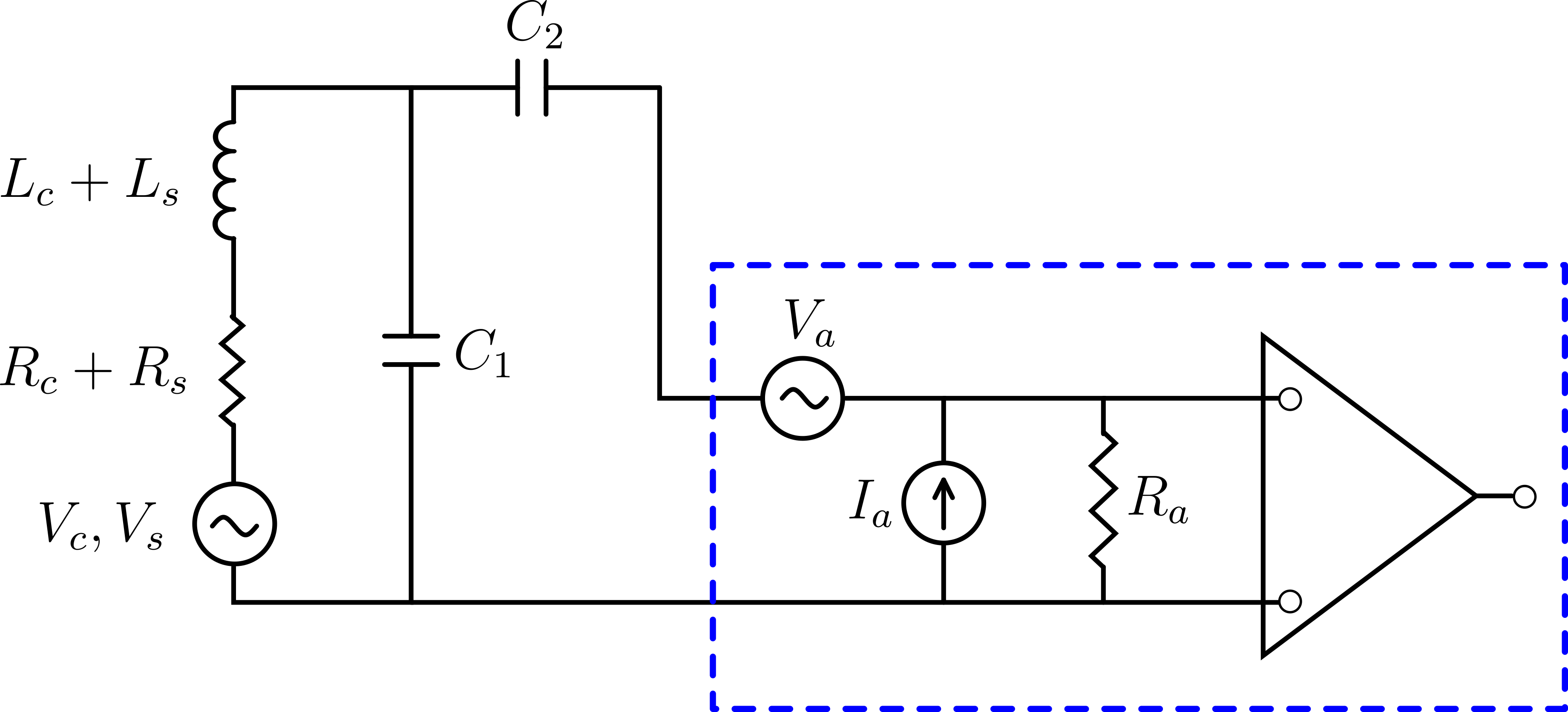}
\end{center}
\newpage
\caption{\linespread{1.0}\fontsize{9}{12}\selectfont
Circuit model of a tuned NMR probe. The pickup coil is shown as an inductor $L_c+L_s$, and the dissipation is shown as a resistor $R_c+R_s$, where $L_s$ and $R_s$ are the contributions due to the spin-ensemble permeability. Capacitors $C_1$ and $C_2$ are tuned so that the circuit resonant frequency is at or near the spin Larmor frequency. The noise sources in this circuit are: spin-projection noise $V_s$, Nyquist noise due to circuit loss $V_c$, and the amplifier, whose noise can be described by the input-referred voltage noise $V_a$ and current noise $I_a$.
}
\label{fig:Probe_circuit_schematic}
\end{figure}
The spin-projection noise is the fundamental limit on sensitivity of magnetic resonance experiments searching for dark matter and other physics beyond the Standard Model. In this section we consider the technical requirements for the category of such experiments, based on the  NMR-probe design of a single-pole LC resonator, capacitively coupled to a voltage amplifier, Fig.\,\ref{fig:Probe_circuit_schematic}.
A number of pickup circuits were analyzed in the NMR literature, e.g. Ref.~\cite{McCoy1989SpinnoiseRT}. Our circuit includes both the probe and the amplifier. This analysis is especially relevant to NMR-based dark matter searches with inductive detection.



\subsection{Noiseless amplifier}

Let us first consider the case of a noiseless amplifier. Then, apart from the spin ensemble, the only source of noise in the circuit is the Nyquist noise in the resistor $R_c$. The power spectral density of this white noise is given by
\begin{align}
\tilde{V}^2_c(\omega) = 2R_ck_B\theta_c/\pi,
\label{eq:405}
\end{align}
if the resistor is at temperature $\theta_c$. 
Since the two noise sources are in series,
we conclude that, in order to reach the spin-projection noise limit, we have to design our experiment so that the spectral density of circuit Nyquist noise is below the spin-projection noise: 
$2R_ck_B\theta_c/\pi<\tilde{V}^2_s(\omega=\omega_0)$. Using Eq.\,(\ref{eq:330}), we can write this requirement as:
\begin{align}
k_B\theta_c<\frac{q}{4}Q_c\mu_0\hbar^2\gamma^2n\omega_0 T_2^*,
\label{eq:410}
\end{align}
where $Q_c=\omega_0L_c/R_c$ is the circuit quality factor without the spin sample present.
Making use of Eqs.~(\ref{eq:M_0}) and (\ref{eq:113}), this condition can also be written as (up to factors of order unity):
\begin{align}
\frac{\theta_s}{\theta_c}\frac{T_2^*}{T_r}>1.
\label{eq:415}
\end{align}
These equations quantify the temperature $\theta_c$ to which the circuit should be cooled in order for spin-projection noise to dominate over the circuit Nyquist noise.


\subsection{Real amplifier, small spin impedance}

When we consider NMR measurements with the circuit shown in Fig.\,\ref{fig:Probe_circuit_schematic}, we have to take into account how the NMR signal and the noise sources are coupled to the amplifier input by the capacitive matching network. Usually capacitors are chosen so that the circuit resonance is at or near the Larmor frequency,
and the impedance is matched to the amplifier input impedance $R_a$. If the presence of the spins does not significantly affect the circuit impedance ($L_s\ll L_c$, $R_s\ll R_c$), and the circuit damping is small ($Q_c\gg 1$), then we can write down the expression for converting the on-resonance voltage amplitude $V_s$ to the voltage $V'_s$ that appears at the input of the amplifier:
\begin{align}
V'_s = \frac{V_s}{2}\sqrt{\frac{Q_cR_a}{\omega_0L_c}}.
\label{eq:420}
\end{align}
Substituting Eq.\,(\ref{eq:340}), we can formulate the amplifier input noise level required to reach the spin-projection noise limit:
\begin{align}
\tilde{V}^2_n(\omega) < Q_cR_a \frac{q}{2\pi}\mu_0\hbar^2\gamma^2n\omega_0 T_2^* .
\label{eq:430}
\end{align}
Note that we have neglected the back-action of the amplifier noise on the spins, but included the circuit back-action via the modified spin coherence time $T_2^*$, see Eq.~(\ref{eq:115}).

\subsection{Real amplifier, large spin impedance}\label{sec:RealAmp}

In order to optimize the sensitivity to beyond-Standard-Model physics, it is often advantageous to work with spin samples with large spin density $n$ and long coherence times $T_2^*$. In such cases the presence of the spin sample leads to large changes in the probe-circuit impedance: $R_s\gtrsim R_c$, $L_s\gtrsim L_c$. This affects signal coupling to the amplifier and can lead to effects such as reduction of noise spectral density near the Larmor frequency, where the spin sample significantly alters the probe circuit impedance. 

We numerically model the noise spectrum that appears at the amplifier input. In our model, the resonant circuit is tuned to $\omega_c=2\pi\times 100\uu{MHz}$, with a circuit quality factor $Q_c=10^3$. The pickup coil has $\eta=4$ turns, 1\,cm length and diameter, and inductance $L_c=0.1\uu{\mu H}$. 
We assume a proton nuclear spin sample with number density $n=10^{23}\uu{cm^{-3}}$ and filling factor $q=0.5$. The nuclear spin Larmor frequency is tuned to $\omega_0 = 2\pi\times 100.1\uu{MHz}$, and the NMR linewidth is 1~ppm in the limit of no radiation damping. We choose a small detuning between the circuit resonance and the Larmor frequency so that the corresponding noise peaks appear at separate frequencies.

We explore the noise spectrum of the system by calculating the different contributions: spin-projection noise, circuit Nyquist noise, and amplifier noise. The spin-projection noise is peaked at the spin Larmor frequency, as described by Eq.\,(\ref{eq:330}). The circuit Nyquist noise is peaked at the circuit resonance frequency, as described by Eq.\,(\ref{eq:405}). The amplifier noise is generated by voltage and current noise sources, with respective spectral densities $\tilde{V}_a^2(\omega)=2k_B\theta_aR_a/\pi$ and $\tilde{I}_a^2(\omega)=2k_B\theta_a/(\pi R_a)$, where $\theta_a$ is the amplifier noise temperature.
The amplifier input impedance $R_a$ and noise impedance $\tilde{V}_a/\tilde{I}_a$ are both set to $R_a=50\uu{\Omega}$.

\begin{figure}[!ht]
\begin{center}
\includegraphics[width=0.9\textwidth]{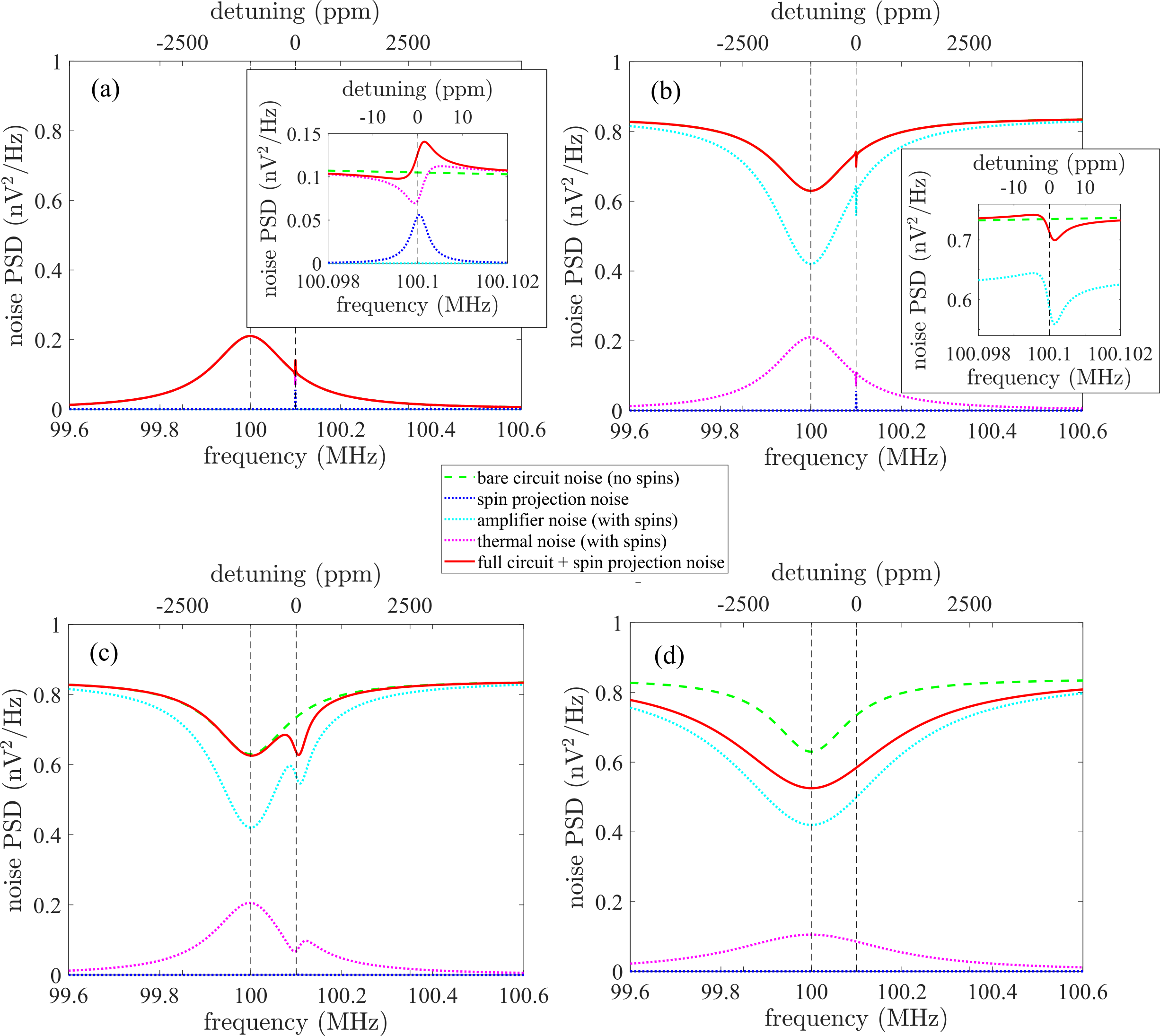}
\end{center}
\newpage
\caption{\linespread{1.0}\fontsize{9}{12}\selectfont
Noise voltage power spectral density (PSD), referred to amplifier input. (a) The case of noiseless amplifier: $\theta_a=0$, $\theta_c=300$~K, $\theta_s=300$~K. (b) The system in thermal equilibrium: $\theta_a=300$~K, $\theta_c=300$~K, $\theta_s=300$~K. (c) Hyperpolarized spin ensemble, moderate circuit back-action:
$\theta_a=300$~K, $\theta_c=300$~K, $\theta_s=3$~K. (d) Hyperpolarized spin ensemble, radiation damping dominates linewidth: $\theta_a=300$~K, $\theta_c=300$~K, $\theta_s=0.03$~K.
The circuit back-action broadens the spin projection noise spectrum and suppresses its peak amplitude for hyperpolarized spin ensembles.
}
\label{fig:2}
\end{figure}

In the noiseless amplifier limit, we observe the effect of the spin ensemble on the pickup probe circuit, Fig.~\ref{fig:2}(a). We note that the circuit Nyquist noise present at the amplifier input is affected by the change in probe circuit impedance due to the spin ensemble (magenta dotted line). It appears that at some frequencies the presence of the spin ensemble decreases the Nyquist noise. This is caused by the change in circuit impedance at those frequencies, due to the spin ensemble. 
The spin-projection noise (blue dotted line) is added to the Nyquist noise, resulting in the solid red line. 

Amplifier noise introduces additional contributions to the noise spectrum, Fig.~\ref{fig:2}(b). The broad dip at the circuit resonance frequency is the feature of our amplifier noise circuit model, which includes independent voltage and current noise sources, added in quadrature. If the pickup probe impedance is $Z_p$, then the total amplifier noise voltage at its inputs is given by
\begin{align}
\tilde{V}_n^2 = \frac{\tilde{V}_a^2R_a^2}{(R_a+Z_p)^2} + \frac{\tilde{I}_a^2R_a^2Z_p^2}{(R_a+Z_p)^2}.
\label{eq:450}
\end{align}
Away from circuit resonance $Z_p\gg R_a$, and $\tilde{V}_n^2 = \tilde{I}_a^2R_a^2 = \tilde{V}_a^2$. But at circuit resonance $Z_p=R_a$, and $\tilde{V}_n^2 = \tilde{V}_a^2/2$. Once again, there are two effects due to the spin ensemble. (1) The frequency-dependent spin impedance $R_s+i\omega L_s$ affects the circuit Nyquist noise and the amplifier noise appearing at the amplifier inputs. (2) The spin-projection noise is added to these noise sources in quadrature. 

Spin hyperpolarization amounts to decreasing the spin temperature $\theta_s$. As discussed in section \ref{sec:SPN}, this does not change the root-mean-squared spin-projection noise, equivalent to the area under its frequency power spectrum. However higher magnetization $M_0$ broadens the NMR line, as a result of the circuit back-action, Eq.~(\ref{eq:115}). Therefore the peak amplitude of the spin-projection noise spectrum decreases. Both effects can be seen in Fig.~\ref{fig:2}(c), for spin temperature $\theta_s=3\uu{K}$, corresponding to (proton) spin polarization of $4\times10^{-4}$. If the spin temperature is another factor of 100 lower, corresponding to 4\% spin polarization, the broadening is so large that it dominates the width of the pickup circuit resonance, Fig.~\ref{fig:2}(d). This broadening due to circuit back-action makes the spin-projection noise more challenging to detect for hyperpolarized spin ensembles.

\clearpage

\section{Spin-projection noise limits for magnetic resonance-based searches for axion-like dark matter}

\noindent
Let us consider the spin-projection noise limits for the CASPEr-electric and CASPEr-gradient experiments, which use nuclear magnetic resonance to search for axion-like dark matter. We will not detail the technical requirements (such as amplifier noise and circuit temperature) necessary to achieve these limits, because doing so would necessitate a detailed optimization of the experimental design parameters, which is beyond the scope of this work. 

Following Eq.~(\ref{eq:140}), we consider the steady-state amplitude of the transverse magnetization, induced by resonant driving of the nuclear spin ensemble: $\mu_0M_1 = \mu_0M_0\Omega_1T_2^*$, where $\Omega_1$ is the drive Rabi frequency, and we assume the drive carrier frequency is close to circuit resonance.
The spin coherence time $T_2^*$ includes the contribution from the probe circuit back-action, given by Eq.~(\ref{eq:113}). 
We convert the transverse magnetization to voltage $V_1$ across the pickup coil using transfer coefficient $\alpha_0$, see section \ref{sec:circuitmodel}. We then compare this signal voltage with the spin-projection noise voltage, given by Eq.~(\ref{eq:330}). After averaging for time $\tau_m$, the smallest detectable signal voltage is:
\begin{align}
V_1^2 = \frac{\tilde{V}_s^2(\omega=\omega_0)}{\sqrt{\tau_m\tau_a}},
\label{eq:463}
\end{align}
where $\tau_a$ is the axion coherence time~\cite{Bud14}. We convert this signal voltage limit to interaction strength to determine the corresponding spin-projection noise limit. This is a naive noise estimate, that neglects the possibility of searching away from the circuit resonance; for a discussion of the optimized search strategy in the context of electromagnetic searches see Ref.~\cite{Chaudhuri2019}.

\subsection{The EDM interaction of axion-like dark matter}

\noindent
The EDM interaction of the axion-like dark matter field $a$ with nuclear spin $I$ is described by the Hamiltonian:
\begin{align}
H_{\rm EDM} = g_da\vec{E}^*\cdot\vec{I}/I,
\label{eq:465}
\end{align}
where $g_d$ is the coupling strength and $E^*$ is an effective electric field~\cite{Bud14}. The search for this interaction, using solid-state NMR, is described in Ref.~\cite{Aybas2021}. For $^{207}$Pb nuclear spins ($I=1/2$) in ferroelectric PMN-PT (chemical formula: \chem{(PbMg_{1/3}Nb_{2/3}O_3)_{2/3} - (PbTiO_3)_{1/3}}) the effective electric field is $E^*=340\uu{kV/cm}$~\cite{Aybas2021}. 

\begin{figure}[!ht]
\begin{center}
\includegraphics[width=0.9\textwidth]{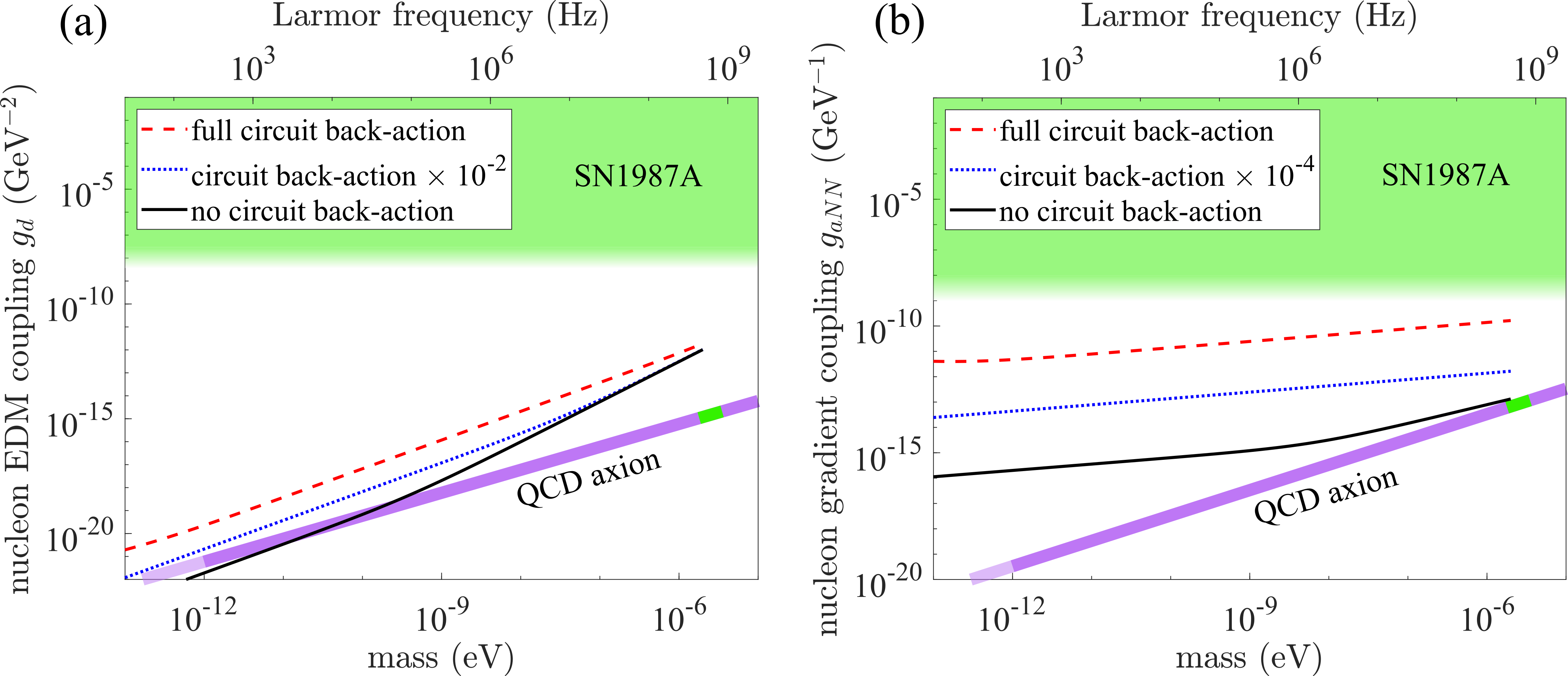}
\end{center}
\newpage
\caption{\linespread{1.0}\fontsize{9}{12}\selectfont
Spin-projection-noise limits for magnetic resonance-based searches for axion-like dark matter. The green region is excluded by analysis of cooling in supernova SN1987A, with color gradient indicating theoretical uncertainty~\cite{Gra13}.
The purple line shows the QCD axion coupling band. The darker purple color shows the mass range motivated by theory~\cite{Gra13}, and the green band marks the mass range where the ADMX experiment is searching for the QCD axion-photon coupling~\cite{Du2018}.
(a) CASPEr-e search for the EDM interaction with a $10\uu{cm}$ sample radius. (b) CASPEr-g search for the gradient interaction with a $10\uu{cm}$ sample radius.
}
\label{fig:4}
\end{figure}

For our estimates we consider a cylindrical volume of ferroelectric PMN-PT with radius $r=10\uu{cm}$ and height equal to diameter. We set the filling factor and the spin polarization to unity, the probe circuit quality factor to $Q_c=10^3$, and the measurement time to $\tau_m=30\uu{min}$. The nuclear spin-coherence time is $T_2=16.7\uu{ms}$ and the chemical shift anistropy is $2000\uu{ppm}$~\cite{Aybas2021}. Even with these relatively short coherence times, circuit back-action (radiation damping) limits the spin coherence and the experimental sensitivity, Fig.~\ref{fig:4}(a). In order to reach the sensitivity at the level of the QCD axion coupling, it is necessary to suppress the circuit back-action by a factor of $\approx10^4$.
As mentioned in section \ref{sec:2.3}, the most promising approach is the pickup-probe feedback scheme.

\subsection{The gradient interaction of axion-like dark matter}

\noindent
The gradient interaction of the axion-like dark matter field $a$ with nuclear spin $I$ is described by the Hamiltonian:
\begin{align}
H_{\rm gr} = g_{\text{aNN}}\vec{\nabla}a\cdot\vec{I},
\label{eq:466}
\end{align}
where $g_{\text{aNN}}$ is the coupling strength~\cite{Bud14}. There have been a number of experimental searches for this interaction, using a variety of spin species~\cite{wu2019search,garcon2019constraints,Abel2017_nEDM,Smorra2019_DM_AM,Bloch2020_comag}. 
For our estimates we consider a cylindrical volume of radius $r=10\uu{cm}$ and height equal to diameter. This volume is filled with proton nuclear spins with number density $n=10^{29}\uu{m^{-3}}$.
We set the filling factor and the spin polarization to unity, the probe circuit quality factor $Q_c=10^3$, and the measurement time $\tau_m=30\uu{min}$. The nuclear spin coherence time is set to $T_2=1\uu{s}$ and the inhomogeneous broadening is $2\uu{ppm}$. 
Circuit back-action (radiation damping) is much more important, given the narrow linewidth assumed for this search, Fig.~\ref{fig:4}(b). In order for back-action not to limit experimental sensitivity, it needs to be suppressed by a factor of $\approx 10^{9}$. 


\section{Driving the spin ensemble -- is there a win?}

The treatment in Sec.\,\ref{sec:RealAmp} shows that the condition that experimental sensitivity be limited by spin-projection noise places stringent requirements on the noise level of the amplifier (see the schematic in Fig.\,\ref{fig:Probe_circuit_schematic}). We consider whether it is possible to manipulate the spin ensemble in such a way as to ease the requirements on amplifier noise. One idea is to ``bias'' the transverse spin signal, by applying an rf pulse, or sequence of pulses, to tilt the magnetization away from the z-axis, prior to the search for a spin torque due to beyond-Standard-Model physics. This would be similar to aligning the sensitive axis of an optical polarimeter at an angle to its dark channel (see, for example, Ref.~\cite{budker2008atomic}, Sec.\,8.9).

Suppose the transverse magnetization due to spin interaction with dark matter is $M_y(t)=M_1\cos(\omega t+\phi)$. We do not know the phase $\phi$, in fact it is random, varying over the dark matter field coherence time (along with the value of $M_1$ \cite{centers2019stochastic}). 
Our apparatus detects voltage $V=\alpha M_y(t)+V_n(t)$, where $\alpha$ is a constant transfer coefficient that depends on apparatus details, and $V_n(t)$ is the noise term. We model the noise voltage as a fluctuating signal with $\langle V_n(t)\rangle = 0$, $\langle V_n^2(t)\rangle = V_n^2$, and $\langle V_n^4(t)\rangle = 3V_n^4$, corresponding to Gaussian noise with variance $V_n^2$. There are three noise contributions: circuit noise, spin noise, and amplifier noise. These are uncorrelated with each other, so to get $V_n^2$ we add their variances. The noise is also uncorrelated with the spin dark matter signal $M_y(t)$.
Over times much longer than dark matter coherence time, $\langle V\rangle = 0$, and we end up having to measure $\langle V^2\rangle = \alpha^2M_1^2/2+V_n^2$,
which in practice means analyzing the voltage power spectral density. The signal-to-noise ratio is: ${\rm SNR} = \alpha^2M_1^2/(2V_n^2)$.

Now let us drive the spin ensemble, for example, by applying a resonant tipping pulse,
so that magnetization is $M_y(t)=M_p\cos(\omega t)+M_1\cos(\omega t+\phi)$, where $M_p$ represents the transverse magnetization that results from the application of the pulse. The tipping angle of the pulse is small but larger than any dark-matter-induced tip, so that $M_p\gg M_1$. The detected voltage is now
\begin{align}
V(t) = \alpha M_p\cos(\omega t)+\alpha M_1\cos(\omega t+\phi)+V_n(t).
\label{eq:433}
\end{align}
The dark matter signal still does not appear in $\langle V\rangle$, but when we square the voltage, there is a cross term $2\alpha^2 M_p M_1\cos(\omega t)\cos(\omega t+\phi)$.
After averaging, this cross term disappears: $\langle V^2\rangle = (\alpha^2M_p^2+\alpha^2M_1^2)/2+V_n^2$,
and, on average, we simply added an offset to our dark matter signal.
However let us calculate the variance of $V^2$:
\begin{align}
\mathrm{var}(V^2) = \langle V^4\rangle - \langle V^2\rangle^2 = \alpha^4M_p^4/8+2V_n^2(\alpha^2M_p^2+V_n^2)+\alpha^2M_1^2(\alpha^2M_p^2+2V_n^2),
\label{eq:450}
\end{align}
where we neglected the term of order $M_1^4$.
Note that we have used the averages $\langle \cos^2{(\omega t)}\rangle = 1/2$, $\langle \cos^4{(\omega t)}\rangle = 3/8$, and $\langle \cos^2{(\omega t)} \cos^2{(\omega t + \phi)} \rangle = 1/4$.
We have ``enhanced'' our dark matter-induced signal by a factor $\alpha^2M_p^2+2V_n^2$. However the signal-to-noise ratio has not improved. For example, in the limit $\alpha M_p\gg V_n$, the dominant noise term is $2\alpha^2M_p^2V_n^2$, and the $\mathrm{SNR}=\alpha^4M_1^2M_p^2/(2\alpha^2M_p^2V_n^2)=\alpha^2M_1^2/(2V_n^2)$, which is independent of $M_p$, and the same as the SNR with $\langle V^2\rangle$ measurement.

Therefore, if the detector response is linear, then there is, in general, no gain in signal-to-noise ratio. In fact we have introduced extra technical complexity, since the pulse magnetization has to be carefully controlled so as not to introduce extra noise into the measurement. However there are certain detection regimes in which multiplying both signal and noise by a common factor can actually lead to an improvement of signal-to-noise ratio at the detector output -- one common example is a photodetector that has a finite dark current. Another possible technical benefit is the loosening of requirements on gain of the first amplifier stage.
The case for which our pulse scheme could offer a substantial advantage is when the dark matter field coherence time is long compared to the measurement time. In this situation, the cross term $2\alpha^2 M_p M_1\cos(\omega t)\cos(\omega t+\phi)$ appearing in the square of the voltage [Eq.\,\eqref{eq:433}] does not average away since the dark matter field phase $\phi$ is constant over the measurement time, and phase cycling should be employed to search for the signal~\cite{garcon2019constraints}. In this regime the SNR could potentially be enhanced by a factor $\sim 2 M_p/M_1$. 



\section{Data-analysis strategy}

\noindent
There are several possible approaches to analyzing the data obtained with a spin-based detector sensitive to new physics. The aim is to extract from the data the maximum amount of information about a particular new physics model. For high signal frequency analysis usually starts by performing the Fourier transformation to convert the data into frequency domain. The computational complexity of fast Fourier transform is $O(N\log{N})$, where $N$ is the number of time-domain data points. In practice, the maximum size of a data block that can be Fourier transformed is often limited by the size of the available memory. It is important, if  possible, to choose the time duration $\tau_b$ of this data block to be longer than the coherence time $\tau_a$ of the new-physics signal that the experiment is searching for. This ensures that the resulting frequency spectrum has $\approx \tau_b/\tau_a$ data points within the signal bandwidth. If the total data-taking time is $\tau_m$, then the power spectral densities of the data blocks (whose total number is $\approx \tau_m/\tau_b$) can be averaged together, in order to improve the signal-to-noise ratio. One way to search for a new-physics signal in the averaged spectrum is by optimal filtering. This is how a number of experiments search for axion-like dark matter (see \cite{Aybas2021} and references therein). For low signal frequencies, the entire experimental run time may be within a single coherence time: $\tau_m<\tau_a$~\cite{wu2019search,garcon2019constraints}. In general, this results in a loss in sensitivity~\cite{centers2019stochastic}. However it is possible to search for coherent signals by using phase-cycling to implement coherent data averaging~\cite{garcon2019constraints}.

If the Gaussian white noise model is a good approximation in the frequency range near a potential signal, then the minimum detectable signal can be estimated as in Eq.~\eqref{eq:463}~\cite{Bud14}. We note that the noise sources considered in this work have some spectral structure. The linewidths due to a circuit resonance are usually much broader than a new-physics signal. We have considered circuit quality factors $Q_c\approx 10^3$, while an axion-like dark matter signal has quality factor of $\approx 10^6$~\cite{Gra13}. However, magnetic resonance linewidths are routinely narrower than $1\uu{ppm}$ and can even be narrower than $1\uu{ppb}$ ~\cite{Bornet2011,nikiel2014ultrasensitive}. 
In this case, the spin-projection noise spectrum can have a linewidth comparable to or narrower than that of a new-physics signal. The optimal data-analysis strategy in this case is yet to be developed.

\section{Outlook}

There is a long road ahead of the CASPEr experiments to reach the spin-projection noise limit. Nevertheless, it is interesting to consider whether this limit could be overcome, at least in principle. We note in this context that some dark-matter experiments (for example, HAYSTAC \cite{Backes2020quantum} and DM Radio \cite{Chaudhuri2018,Chaudhuri2019b}, 
where the claim is that the use of squeezing, entanglement, and back-action evading techniques leads to a sensitivity better than the standard quantum limits. 
Perhaps, the most promising aspect of such quantum techniques is that they may improve the measurement bandwidth \cite{auzinsh2004can}.


\section{Acknowledgements}

The authors acknowledge valuable discussions with Stephen Kuenstner and Kent Irwin.
The authors at BU acknowledge support from US Department of Energy grant DESC0019450, the Heising-Simons Foundation grant 2015-039, the Simons Foundation grant 641332, National Science Foundation under grant PHY-1806557, and the Alfred P. Sloan foundation grant FG-2016-6728.
The work of the Mainz group was supported in part by the Cluster of Excellence PRISMA+ funded by the German Research Foundation (DFG) within the German Excellence Strategy (Project ID 39083149), by the European Research Council (ERC) under the European Union Horizon 2020 research and innovation program (project Dark-OST, grant agreement No 695405), and by the DFG Reinhart Koselleck project. AW acknowledges support from the German Federal Ministry of Education and Research (BMBF) within the Quantumtechnologien program (FKZ 13N15064).
DFJK acknowledges the support of the National Science Foundation under grant PHY-1707875.

\printbibliography

\end{document}